\documentstyle[12pt]{article}
\begin{document}
\setcounter{page}{1}

{\hfill Yadernaya Fizika, 1998, vol.61, No.10, p.1884-1888}

{\hfill
\vspace{0.5cm}
\begin{center}
 {\bf
EXACT SOLUTIONS OF NONSTATIONARY \\SHR\"ODINGER EQUATIONS
and GEOMETRIC PHASE}
\vspace{0.5cm}

{\large \sf A.A. Suzko}
\footnote{
Radiation Physics
and Chemistry Problems Institute, Academy of Sciences of Belarus, Minsk},
 {\large \sf E.P. Velicheva}
\footnote{
Gomel State University, Belarus}\\
\vspace{0.5cm}

{\it Joint Institute for Nuclear Research, Dubna, Russia}
\end{center}
\vspace{0.5cm}

{\bf Abstract.}
The procedure of solving nonstationary Schr\"odinger equations
in the exact analytic form is elaborated on the basis of exactly solvable
stationary models. The exact solutions are employed to study
the nonadiabatic geometric phase at cyclic evolution of quantum systems.

\section{Introduction}
Consider the time-dependent Schr\"odinger equation
\begin{eqnarray}
\label{1.1}
i\hbar\frac{\partial |\Psi(t)>}{\partial t} = H(t)|\Psi(t)>
\end{eqnarray}
the exact solution of which is not known in general.
The purpose of this paper is to give the procedure of obtaining a wide
class of Hamiltonians for which the exact solutions of (\ref{1.1})
can be found.

Introduce a unitary transformation
$ {\cal S}^{\dagger}(t) = {\cal S}^{-1}(t)$
\begin{eqnarray}
\label{1.3}
|\Psi(t)> = {\cal S}(t)|\eta(t)>
\end{eqnarray}
such that in the new equation
\begin{eqnarray}
\label{1.1a}
i\hbar\frac{\partial |\partial\eta(t)>}{dt} = \bar{H}|\eta(t)>
\end{eqnarray}
the Hamiltonian
\begin{eqnarray}
\label{1.4}
\bar{H}={\cal S}^{\dagger}(t)H(t){\cal S}(t)-
i\hbar {\cal S}^{\dagger}(t) \dot{\cal S}(t)~
\end{eqnarray}
does not depend on time, i.e., $(\partial / \partial t)\bar{H}=0$
~\cite{Wang,Wu}.
From this, we obtain the equation for the unitary transformation {\cal S}(t)
\begin{eqnarray}
\label{1.8}
\dot{{\cal S}^{\dagger}}H(t){\cal S}+{\cal S}^{\dagger}(\partial{H}(t)/\partial t){\cal S}+
{\cal S}^{\dagger}H(t)\dot{\cal S}
-i\hbar\dot{{\cal S}^{\dagger}} \dot{{\cal S}}
-i\hbar{\cal S}^{\dagger} \ddot{\cal S}=0,
\end{eqnarray}
where
$\dot{\cal S}=(d/dt){\cal S} $.

From (\ref{1.4}) for $t=0$ it follows
\begin{eqnarray}
\label{1.4a}
\bar{H}={\cal S}^{\dagger}(0)H_o{\cal S}(0)-
i\hbar {\cal S}^{\dagger}(0) \dot{\cal S}(0).
\end{eqnarray}
Taking account of this equation in (\ref{1.4}) we obtain
\begin{eqnarray}
\label{1.4b}
H(t)={\cal S}(t){\cal S}^{\dagger}(0)H_o{\cal S}(0){\cal S}^{\dagger}(t),
\end{eqnarray}
where $H_o=H(t=0)$.

Let us define the solution to equation (\ref{1.1a}).
On the one hand, it is
\begin{eqnarray}
\label{1.5a}
\eta(t)= \exp(-i\bar{H}t/\hbar)\eta(0),
\end{eqnarray}
on the ather hand, it is
\begin{eqnarray}
\label{1.6a}
 \eta(t)=\exp(-i\bar{H}t/\hbar )\Phi({\cal E}) \nonumber.
\end{eqnarray}
Therefore, $\eta(0)=\Phi({\cal E})$ is the function of the initial state
of (\ref{1.1a}) and is one of the eigenfunctions of $\bar{H}$
\begin{eqnarray}
\label{1.13a}
 \bar{H}\Phi({\cal E})={\cal E}\Phi({\cal E}).
\end{eqnarray}

Now the solution (\ref{1.1}) is reduced to the solutions of equations
(\ref{1.1a}) and (\ref{1.8}). The solution of the equation (\ref{1.1a})
becomes the eigenequation problem (\ref{1.13a}) for the time-independent
Hamiltonian $\bar{H}$. But solving (\ref{1.8}) is in general a very
complex problem. On the other hand, it is possible to assume the concrete
forms of ${\cal S}(t)$ and seek the Hamiltonian $H(t)$ that satisfies
(\ref{1.8}). Then, to obtain exact solutions of (\ref{1.1}) we, in fact,
need an exactly solvable time-independent equation (\ref{1.13a}).
After this, by using the concrete forms of ${\cal S}(t)$, we can construct
a wide class of exactly solvable time-dependent $H(t)$ ~\cite{report}.

In terms of the time-evolution operator ${\cal U}(t)={\cal U}(t,0)$
defined by, $\Psi(t)={\cal U}(t)\Psi(0)$,  the Schr\"odinger equation
(\ref{1.1}) is written as
\begin{eqnarray}
\label{1.2}
i\hbar\dot{\cal U}(t) = H(t){\cal U}(t).
\end{eqnarray}
It is not difficult to derive a very important relationship between
the operators ${\cal U}(t)$ and ${\cal S}(t)$
\begin{eqnarray}
\label{1.U}
{\cal U}(t)={\cal S}(t)\exp(-i{\bar H}t/\hbar){\cal S}^{\dagger}(0)
\end{eqnarray}
because if ${\bar H}$ and ${\cal S}(t)$ are known,
${\cal U}(t)$ is known, too.
\section{ Construction of the Hamiltonian of the
nonstationary Schr\"odinger equations}
Consider an example of reconstruction of a time-dependent Hamiltonian
\begin{eqnarray}
\label{1.18a}
H(t)=({\hat p}_{x}^{2}+ q(x))\hat I +{\bf B}(t,x)\cdot{\bf j}
\end{eqnarray}
for which equation (\ref{1.1}) is exactly solved. Here
${\hat p}_{x}=-i\hbar\nabla_{x}$ is the operator of momentum, $q(x)$
is a stationary potential and ${\bf B}(t,x)$ is a time-varying potential
that can be considered as an analog of the inhomogeneous magnetic field,
${\bf j}= ({\hat j_{x}},{\hat j_{y}},{\hat j_{z}})$ is
the spin operator
and ${\hat j_{i}}=(2{j}+1)\times(2{j}+1)$ are matrices.

Let the model Hamiltonian $\bar{H}(x)$ of the stationary problem correspond
to two-channel one
\begin{eqnarray}
\label{1.15a}
\bar{H}(x)&=&{\hat p}_{x}^{2}
\left( \matrix{1 & 0  \cr
	       0 & 1 \cr}\right)+
\left( \matrix{V_{11}(x) & V_{12}(x)  \cr
	       V_{21}(x) & V_{22}(x) \cr}\right)
=\Bigl( {\hat p}_{x}^{2}
+\frac{V_{11}(x)+V_{22}(x)}{2}\Bigr)
\left( \matrix{1 & 0  \cr
	       0 & 1 \cr}\right) \nonumber \\
&+&\left( \matrix{(V_{11}(x)-V_{22}(x))/2 & V_{12}(x)  \cr
	       V_{21}(x) & -(V_{11}(x)-V_{22}(x))/2 \cr}\right).
\end{eqnarray}
It can be rewritten as
\begin{eqnarray}
\label{1.16}
\bar{H}(x)=(\hat{p}_{x}^{2} +q(x))\hat I+ \bar{\bf B}(x) \cdot \mbox{\boldmath$\sigma$}
\end{eqnarray}
$$=({\hat p}_{x}^2 +q(x))\hat I +
\bar{\Omega}(x)\left ( \matrix{ \cos\bar{\theta}(x) & \sin\bar{\theta}(x)  \cr
	        \sin\bar{\theta}(x) & -\cos\bar{\theta}(x) \cr}\right)$$
with the evident notation:
$ q(x)=(V_{11}(x)+V_{22}(x))/2$, $\hat{I}$ is the unit matrix,
$\mbox{\boldmath$\sigma$}=(\hat{\sigma}_1,\hat{\sigma}_2,\hat{\sigma}_3)$,
$\bar{\bf B}(x)$ is expressed in terms of $V_{ij}(x)$ and can be defined as
\begin{eqnarray}
\label{1.17}
\bar{\bf B}(x)= \bar{\Omega}(x)( \sin\bar{\theta}(x), 0, \cos\bar{\theta}(x)),
\end{eqnarray}
$$\bar{\Omega}(x) = 1/2\sqrt{(V_{11}(x)-V_{22}(x))^{2} + 4V_{12}(x)^{2}},$$
$$\sin\bar{\theta}(x)= \frac {V_{12}(x)} {\bar{\Omega}(x)},~
\cos\bar{\theta}(x)=\frac{V_{11}(x)-V_{22}(x)}{2\bar{\Omega}(x)}.$$
By means of a canonical unitary transformation ${\cal S}(t)$, one can
obtain a time-dependent Hamiltonian $H(t)$ in an explicit form
 from the stationary Hamiltonian (\ref{1.16}).
If ${\cal S}(t)$ is chosen as an operator of rotation around $z$-axis
\begin{eqnarray}
\label{S(t)}
{\cal S}(t)=\exp(-i{\hat \sigma}_3\omega t),
\end{eqnarray}
then in accordance with (\ref{1.4}) the following relationship
between the stationary ${\bar{H}}$ and nonstationary $H(t)$ Hamiltonians is
obtained
\begin{eqnarray}
\label{1.44}
H(t)={\cal S}(t){\bar H}(x){\cal S}^{\dagger}(t)+\omega\hat{\sigma}_{3}.
\end{eqnarray}
This time-dependent Hamiltonian with an allowance for (\ref{1.16}) can
be rewritten as
\begin{eqnarray}
\label{1.16c}
H(t)= ((\hat{p}_{x}^{2} +q(x))\hat{I}
+\bar{\Omega}(x)\left ( \matrix{ \cos\bar{\theta}(x)+\omega/\bar{\Omega}(x) &
                                \sin\bar{\theta}(x)\exp(-2i\omega t)  \cr
                                \sin\bar{\theta}(x)\exp(+2i\omega t) &
                              - (\cos\bar{\theta}(x)+\omega/\bar{\Omega}(x)) \cr}\right).
\end{eqnarray}
This may correspond to the problem of two-level atom for the spin-1/2
representation of ${\bf j}$.
Obviously, it corresponds to the Hamiltonian (\ref{1.18a}) with
the nonstationary field
\begin{eqnarray}
\label{1.19}
{\bf B}(t,x)
={\Omega}(x)(\sin{\theta}(x)\cos(2\omega t),
\sin{\theta}(x)\sin(2\omega t),\cos{\theta}(x)),
\end{eqnarray}
the components of which are expressed in terms of the components of the
time-independent field
\begin{eqnarray}
\label{1.20}
\Omega(x) =\bar{\Omega}(x)\Bigl(1+{\omega^2\over\bar{\Omega}^2(x)}+
2{\omega\over\bar{\Omega}(x)}\cos\bar{\theta}(x)\Bigr)^{1/2},
\end{eqnarray}
$$\sin\theta(x)=\frac{\bar{\Omega}(x)\sin\bar{\theta}(x)}{\Omega(x)},
~\cos\theta(x)=
\frac{\bar{\Omega}(x)\cos\bar{\theta}(x)}{\Omega(x)}+\frac{\omega}{\Omega(x)}.$$
As a result of the unitary transformation $S(t)$, the $B_{2}(t,x)$-
component of the vector ${\bf B}(t,x)$ arises.
The expression for the
initial Hamiltonian $ H_{o}(x) $ is directly obtained from (\ref{1.44})
or from (\ref{1.16c})  at $t=0$
\begin{eqnarray}
\label{Ho}
 H_{o}(x)={\bar H}(x)+ \omega \hat{\sigma}_{3}.
\end{eqnarray}
This relation is conveniently rewritten in the form
\begin{eqnarray}
\label{1.o}
 H_{o}(x)= (\hat{p}_{x}^2+q(x))\hat{I} +
{\bf B}_{o}(x)\cdot\mbox{\boldmath$\sigma$},
\end{eqnarray}
where the potential ${\bf B}_{o}(x)$ coincides with ${\bf B}(t,x)$
defined in (\ref{1.19}) at the initial moment $t=0$
\begin{eqnarray}
\label{1.17a}
{\bf B}_{o}(x)=
\Omega(x)( \sin\theta(x), 0, \cos\theta(x)).
\end{eqnarray}
In accordance with (\ref{1.44}) and (\ref{Ho}) and with allowance of
(\ref{S(t)}) the time-varying Hamiltonian  $ H(t)$ is written in terms of the
initial Hamiltonian as
\begin{eqnarray}
\label{H.3}
H(t,x)=
 \exp(-i{\hat \sigma}_3\omega t)H_o(x)\exp(+i{\hat \sigma}_3\omega t)=
 (\hat{p}_{x}^2+q(x))\hat{I}+ {\bf B}(t,x) \cdot \mbox{\boldmath$\sigma$}.
\end{eqnarray}
It means that the time-dependent part of the reconstructed potential is
in precession along the $z$-axis with the frequency $ 2\omega$.
Thus, one has been built the time-dependent Hamiltonian
\begin{eqnarray}
\label{1.19a}
H(t)&=&\Bigl( {\hat p}_{x}^{2}+\frac{V_{11}(x)+V_{22}(x)}{2}
\Bigr)\hat{I} \nonumber\\
&+&
\left(\matrix{\frac{V_{11}(x)-V_{22}(x)}{2}+\omega &
V_{12}(x)\exp(-2i\omega t) \cr
V_{21}(x)\exp(+2i\omega t) & -
(\frac{V_{11}(x)-V_{22}(x)}{2}+\omega)\cr}\right)
\end{eqnarray}
that admits  exact solutions of the nonstationary Schr\"odinger equations
(\ref{1.1}) if the corresponding canonical stationary equation (\ref{1.13a})
with the Hamiltonian  (\ref{1.15a}) is exactly solved
\begin{eqnarray}
\label{1.Psi}
|\Psi(t,x)>=\exp(-i{\hat \sigma}_{3}\omega t)\exp(-i{\bar
H}t/\hbar)|\Phi(x)>.
\end{eqnarray}
Indeed, the eigenvalues ${\cal E}$ of the stationary equation with
the Hamiltonian ${\bar H}$ are known because they are initial data
for the inverse scattering problem.
By using the algebraic procedure of the Bargmann-Darboux transformations,
the potentials $V_{\alpha\alpha'}(x)$ in (\ref{1.15a}) and solutions
$|\Phi(x)>$ are reconstructed in the exact analytic form from the known
spectral data.

Let $V(x)$ be a Bargmann potential matrix for which the system of
equations (\ref{1.13a}) has exact solutions.
For example, matrix elements of the transparent potential
and pertinent Jost solutions are \cite{book}
\begin{eqnarray}
\label{1.V}
V_{\alpha\alpha'}(x)=2\frac{d}{dx}\sum_{\nu \lambda}\exp(-\kappa^{\nu}_{\alpha}x)
\gamma_{\alpha}^{\nu} P_{\nu\lambda}^{-1}(x)
\gamma_{\alpha'}^{\lambda}\exp(-\kappa^{\lambda}_{\alpha'}x),
\end{eqnarray}
\begin{eqnarray}
\label{1.F}
F_{\alpha\alpha'}^{\pm}(k,x)&=&\exp(\pm ik_{\alpha}x)\delta_{\alpha\alpha'} \\ \nonumber
&-& \sum_{\nu \lambda}\gamma_{\alpha}^{\nu}
\exp(-\kappa_{\alpha}^{\nu}x)P_{\nu\lambda}^{-1}(x)\int_{x}^{\infty}
\gamma_{\alpha'}^{\lambda}\exp(-(\kappa_{\alpha'}^{\lambda}\pm ik_{\alpha'})x')dx',
\end{eqnarray}
where
$$P_{\nu\lambda}=\delta_{\nu\lambda}+\sum_{\alpha'}^{m}
\frac{\gamma_{\alpha'}^{\nu} \gamma_{\alpha'}^{\lambda}}
{\kappa^{\nu}_{\alpha'}+\kappa^{\lambda}_{\alpha'}}
 \exp(-(\kappa^{\nu}_{\alpha'}+\kappa^{\lambda}_{\alpha'})x). $$
The normalized solution for the vector function of the bound state is
immediately expressed through the matrix Jost solutions taken
at the energy of the bound state
$k_{\nu}^{2}=-{\cal E}_{\nu},~k_{\alpha}^{\nu}=k_{\nu}\hat{I}$
$$|\Phi_{\alpha}({\cal E}_{\nu},x)>=\sum_{\alpha'}^{m}F_{\alpha\alpha'}({\cal E}_{\nu},x)
\gamma_{\alpha'}^{\nu},~~~\alpha,\alpha'=1,2. $$
Here, the indices $\nu$ and $\lambda$ correspond to the bound states
characterized by the energies ${\cal E}_{\nu}$ and normalizing
matrices $|\gamma^{\nu}_{\alpha}><\gamma^{\nu}_{\alpha'}|$ and the channel
indices are labeled as $\alpha,\alpha'$.
It is worth noting that the class of exactly solvable stationary problems is
very wide. Algebraic  Bargmann-Darboux transformations give multiple
examples of such problems that can be used to obtain the exact solution
of the Schr\"odinger equations for a large class of explicit
time-dependent Hamiltonians.

\section{ The geometric phase}
By means of the unitary transformations $\bar{\cal S}(x)$
and ${\cal S}(x)$ one can annihilate the $\bar{B}_1(x)$-- and
$B_1(x)$--components of the fields $\bar{{\bf B}}(x)$ and ${\bf B}_{o}(x)$,
respectively.
It is realized by means of the $SU(2)$--operators
\begin{eqnarray} \label{SU.2}
\bar{\cal S}(x) =\exp(-i\bar{\theta}(x)\hat{j}_{2})
\mbox~~and ~~\mbox {\cal S}(x)=\exp(-i{\theta}(x)\hat{j}_{2}).
\end{eqnarray}
%
%
Since the following relations are valid:
 \begin{eqnarray} \label{3.2}
\bar{{\bf B}}(x)\cdot\mbox{\boldmath$\sigma$}=\bar{\Omega}(x)\exp(-i\bar{\theta}(x){\hat j}_{2})
{\hat \sigma}_{3}\exp(i\bar{\theta}(x){\hat j}_{2}),
\end{eqnarray}
\begin{eqnarray}
\label{3.1}
{\bf B}_{o}(x)\cdot\mbox{\boldmath$\sigma$}=\Omega(x)\exp(-i{\theta}(x){\hat
j}_{2}){\hat \sigma}_{3} exp(i{\theta}(x){\hat j}_{2}),
\end{eqnarray}
under these transformations $\bar{\cal S}(x)$ and ${\cal S}(x)$,
the systems of equations with the Hamiltonians (\ref{1.16}) and (\ref{1.o})
are reduced to the gauge-type equations \cite{Suzko}
\begin{eqnarray}
\label{3.3}
 [- (\nabla_{x}\hat{I}+\bar{A}(x))^2 + q(x)\hat{I} +
 \bar{\Omega}(x)\cdot\hat{\sigma}_{3} - P^{2}] \Phi'(x) = 0,
 \end{eqnarray}
\begin{eqnarray}
\label{3.4}
 [- (\nabla_{x}\hat{I}+A(x))^2 + q(x)\hat{I} + \Omega(x)\cdot\hat{\sigma}_{3}
- P^{2}] \Psi'_{o}(x) = 0, \\
			  P=diag(p_{\alpha}) \nonumber
\end{eqnarray}
for the new functions $\Phi'(x)$ and $\Psi'_{o}(x)$
\begin{eqnarray}
\label{3.7}
\Phi'(x) = \exp(-i\bar{\theta}(x){\hat j}_{2})\Phi(x),~~~
\Psi'_{o}(x) = \exp(-i{\theta}(x){\hat j}_{2})\Psi_{o}(x).
\end{eqnarray}
The matrix elements of the effective vector potentials $ \bar{A}(x) $ and
$A(x)$ in (\ref{3.3}) and (\ref{3.4}) are generated by the procedure of the
gauge transformation (\ref{SU.2})
\begin{eqnarray}
\label{3.6}
\bar{A}(x)&=& \exp(i\bar{\theta}(x){\hat j}_{2})(d/dx)
\exp(-i\bar{\theta}(x){\hat j}_{2}),\\
A(x)&=& \exp(i{\theta}(x){\hat j}_{2})(d/dx)
\exp(-i{\theta}(x){\hat j}_{2}).\nonumber
\end{eqnarray}
Note, the elements of $\bar{\cal S}(x)$
$$\bar{\cal S}(x)=\exp(-i\bar{\theta}(x){\hat j}_{2})=
\left ( \matrix{ \cos\bar{\theta}(x)/2 &
                                \sin\bar{\theta}(x)/2 \cr
                               - \sin\bar{\theta}(x)/2 &
                             \cos\bar{\theta}(x)/2 \cr}\right)
$$
are defined by the elements of the known potential matrix $V_{\alpha\alpha'}(x)$
(\ref{1.17}).
Taking into account (\ref{S(t)}) for the transformation operator
${\cal S}(t)$ in (\ref{1.U}), we get
$${\cal U}(t)=\exp(-i{\hat \sigma}_{3}\omega t)\exp(-i{\bar H}t/\hbar).$$
Then, the solution (\ref{1.Psi}) of (\ref{1.1}) with the Hamiltonian
(\ref{1.19a}) can be rewritten as
 $$ |\Psi_{\alpha}(t,{\cal E}_{\nu})>=\exp(\mp i\omega t)\exp(-i{\cal E}_{\nu}t/\hbar)
|\Psi_{\alpha}(0,{\cal E}_{\nu})>.$$
The eigenstates and eigenvalues of
 ${\bar H}(x)$ are
$|\Phi_{\alpha}({\cal E}_{\nu})>$ and ${\cal E}_{\nu}=
{\cal E}_{\nu}^{\alpha}\hat{I}$, $\alpha=1,2$. After one period
 $ T=\pi/\omega$, the solution is
\begin{eqnarray}
\label{3.9}
|\Psi_{\alpha}(T,{\cal E}_{\nu})>=\exp(\mp i\pi)\exp(-i{\cal E}_{\nu}T/\hbar)
|\Psi_{\alpha}(0,{\cal E}_{\nu})>.
\end{eqnarray}
and the total phase is
\begin{eqnarray}
\label{3.10}
\delta_{\nu}=(\pm \pi+{\cal E}_{\nu}T/\hbar).
\end{eqnarray}
Next, to find the dynamical phase we calculate the expectation value
of $H(t)$ by using the relations (\ref{1.44}) and (\ref{1.3}) or
(\ref{1.Psi})
$$<\Psi_{\alpha}(t,{\cal E}_{\nu})|H(t)|\Psi_{\alpha}(t,{\cal E}_{\nu})>=
<\Phi_{\alpha}({\cal E}_{\nu}){\cal S}^{\dagger}(t)|H(t)|{\cal S}(t)
\Phi_{\alpha}({\cal E}_{\nu})>$$
\begin{eqnarray}
\label{3.dyn}
=<\Phi_{\alpha}({\cal E}_{\nu})|\bar{H}(x)|\Phi_{\alpha}({\cal E}_{\nu})>
+<\Phi_{\alpha}({\cal E}_{\nu})|\omega \hat{\sigma}_3
|\Phi_{\alpha}({\cal E}_{\nu})>={\cal
E}_{\nu}+\omega\bar{\hat{\sigma}}_3^{\nu}.
\end{eqnarray}
By virtue of (\ref{3.7}) for the eigenfunctions of
$ \bar{H}(x)$ the spin-expectation value  $\bar{\hat{\sigma}}_3^{\nu}$
is written as
\begin{eqnarray}
\label{3.10a}
\bar{\hat{\sigma}}_3^{\nu}=<\Psi_{\alpha}(t,{\cal E}_{\nu})|
\hat{\sigma}_3 |\Psi_{\alpha}(t,{\cal E}_{\nu})>=
<\Phi_{\alpha}({\cal E}_{\nu})|\hat{\sigma}_3 |\Phi_{\alpha}({\cal E}_{\nu})>
\end{eqnarray}
$$=<\Phi'_{\alpha}({\cal E}_{\nu})|\exp(i\bar{\theta}(x){\hat j}_{2})
\hat{\sigma}_{3}
\exp(-i\bar{\theta}(x){\hat j}_{2} |\Phi'_{\alpha}({\cal E}_{\nu})> $$
\begin{eqnarray}
\label{3.10b}
=<\Phi'_{\alpha}({\cal E}_{\nu})|-\sin(\bar{\theta}(x)){\hat \sigma}_{1}
+\cos(\bar{\theta}(x))\hat{\sigma}_{3}|\Phi'_{\alpha}({\cal E}_{\nu})>.
\end{eqnarray}
Then, for the dynamical phase  $\delta_{\nu}^{d} $ we get the expression
\begin{eqnarray}
\label{3.11}
&&\delta_{\nu}^{d}=\int\limits_{0}^{T}<\Psi_{\alpha}(t)
|H(t)| \Psi_{\alpha}(t)>_{\nu}dt \nonumber\\
&&={\cal E}_{\nu}T/\hbar+\pi<\Phi'_{\alpha}({\cal
E}_{\nu})|-\sin(\bar{\theta}(x)) {\hat\sigma}_{1}+
\cos(\bar{\theta}(x))\hat{\sigma}_{3})|\Phi'_{\alpha}({\cal E}_{\nu})>.
\end{eqnarray}
Removing the dynamical phase (\ref{3.11}) from the total phase
(\ref{3.10}) we get the relation for the nonadiabatic geometric phase
$\delta_{\nu}^{g}$
\begin{eqnarray}
\label{3.12}
\delta_{\nu}^{g}=(\delta_{\nu}-\delta_{\nu}^{d})
&=&\pi<\Phi'_{\alpha}({\cal E}_{\nu})|\sin(\bar{\theta}(x)){\hat\sigma}_{1}
+[\pm 1-\cos(\bar{\theta}(x))\hat{\sigma}_{3})]
|\Phi'_{\alpha}({\cal E}_{\nu})> \nonumber \\
&=&\pi(\pm 1 - \bar{\hat{\sigma}}_3^{\nu}).
\end{eqnarray}
One can easily see that the geometric phase is determined by
the spin-expectation value  $\bar{\hat{\sigma}}_3^{\nu}$ along the
rotating axis.
For the first time this result was obtained by Wang in \cite{Wang}
for a spin particle in a rotating magnetic field. As distinct from
\cite{Wang} this approach is based on the exact solutions
$|\Phi_{\alpha}(x)>$ of the body-fixed Hamiltonian (\ref{1.16}).
In the case when the dependence on the space variable is absent,
our problem is simplified and goes to  the problem for the
spin-$1/2$ particle in the rotating magnetic field \cite{Stone}
with the cranking Hamiltonian
\begin{eqnarray}
\label{h.3}
H(t)= \exp(-i\hat{\sigma}_{3}\omega t)H_o\exp(+i\hat{\sigma}_{3}\omega t)=
{\bf B}(t) \cdot \mbox{\boldmath$\sigma$},
\end{eqnarray}
$${\bf B}(t)= {\Omega}( \sin{\theta}\cos(2\omega t),
\sin{\theta}\sin(2\omega t), \cos{\theta})$$
instead of the inhomogeneous $H(t,x)$ in (\ref{H.3}) or (\ref{1.16c}).
It is evident from (\ref{Ho}) that the Routhian (the body-fixed Hamiltonian
$\bar H(x)$, (\ref{1.16}) ) turns into
\begin{eqnarray}
\label{h.2}
\bar{H}=H_o-\omega\hat{\sigma}_{3}= \bar {\bf B} \cdot \mbox{\boldmath$\sigma$}
\end{eqnarray}
with the renormalized magnetic field
$${\bar{\bf B}}={\bar\Omega}(\sin{\bar\theta}, 0, \cos{\bar \theta}).$$
The initial Hamiltonian $H_o$ of this model is
\begin{eqnarray}
\label{h.1}
H_{o}={\bf B}_o \cdot  \mbox{\boldmath$\sigma$},~~~~
{\bf B}_{o}= \Omega( \sin\theta, 0, \cos\theta).
\end{eqnarray}
The total phase is determined as in (\ref{3.10})
and the dynamical phase can be obtained from (\ref{3.dyn}) taking
account of (\ref{h.3}) for $H(t)$ and
with the spin alignment $ \bar{\hat{\sigma}}_3^{m}$ determined as
\begin{eqnarray}
\label{4.10a}
\bar{\hat{\sigma}}_3^{m}=<\Psi_{m}(t)|
\hat{\sigma}_3 |\Psi_{m}(t)>=
<\Phi_{j m}|\hat{\sigma}_3 |\Phi_{j m}>
\end{eqnarray}
$$=<j m|\exp(i\bar{\theta}{\hat j}_{2})
\hat{\sigma}_{3}
\exp(-i\bar{\theta}{\hat j}_{2} |j m> $$
\begin{eqnarray}
\label{4.10b}
=<j m|-\sin(\bar{\theta}){\hat \sigma}_{1}
+\cos(\bar{\theta})\hat{\sigma}_{3}|j m> =m\cos(\bar{\theta}),
\end{eqnarray}
where $ |j m> $ are eigenfunctions of $ \hat{\sigma}_3$, the channel index is
$m$ instead of $\alpha$.
Thus, the dynamical phase is
$$\delta_{m}^{d}={\cal E}_{m}T/\hbar + m\pi\cos(\bar{\theta})$$
and the geometric phase is
\begin{eqnarray}
\label{4.g}
\delta_{m}^{g}=m\pi(1-\cos(\bar{\theta}))
\end{eqnarray}
In the adiabatic limit, when $\omega/\Omega \to 0$, then
$\Omega \to \bar{\Omega},\cos\theta \to \cos\bar{\theta}$
and the geometric phase (\ref{4.g}) becomes Berry's phase \cite{Berry}
$$ \delta_{m}^{g}=m\pi(1-\cos\theta).$$
Note, the relation (\ref{3.10a}) corresponds to the nonadiabatic
Aharonov--Anandan geometric phase \cite{aharon}
\begin{eqnarray}
\label{3.13}
\delta^{g}=\int\limits_{0}^{T}dt<\Psi'(t)|(d/dt) \Psi'(t)>=
\int\limits_{0}^{T}dt<\Phi_{\alpha}(x)|{\cal S}^{\dagger}(t)\dot{\cal S}(t)|
\Phi_{\alpha}(x)>=T\omega\bar{\hat{\sigma}}_3=\pi\bar{\hat{\sigma}}_3.
\end{eqnarray}
We have used the stationary two-channel exactly solvable models and a quite
simple time-dependent unitary operator ${\cal S}(t)$ of rotation to
construct the time-dependent Hamiltonians and the corresponding solutions
of the nonstationary Schr\"odinger equations in a closed analytic form.
It is evident, how the suggested approach is generalized for another choice
of stationary models and time-dependent ${\cal S}(t)$-operators.

\section{Conclusion}

The procedure is given of constructing time-dependent Hamiltonians
for which nonstationary Schr\"odinger equations have exact solutions.
Time-dependent unitary transformations and exactly solvable models
for stationary Schr\"odinger equations are used to obtain
a large class of explicit time-dependent Hamiltonians.
 The two-channel problem has been  considered in detail,
and exact solutions are used to calculate the nonadiabatic geometric
phase.

{}


\begin{thebibliography}{}
\bibitem{Wang}
Wang, Shun-Jin, {\sl Phys. Rev.A}, 1990, vol. 42, p. 5107; ibid p. 5103.
\bibitem{Wu}
Wu, Lian-Ao, Sun, J. and Zhong, Ji-Yu,
{\sl Phys. Lett.A}, 1993, vol. 183, p. 257.
\bibitem{report}
Suzko, A.A. and Velicheva, E.P., Theor.Math.Phys., 1998, vol.115, No.3,
p.410; {\sl JINR Preprint 4-97-311 }, 1997, Dubna, 9p.
\bibitem{book}
 Zakhariev, B.N. and Suzko, A.A.,
{\sl Direct and inverse problems}, (Potentials in quantum scattering),
Springer-Verlag. Berlin Heidelberg/New York, 1990, 223p. 2-nd ed.
\bibitem{Suzko}
Suzko, A.A., {\sl
Phys. Part. Nucl.}, 1993, vol. 24, p. 485.
\bibitem{Stone}
Stone, M., {\sl Phys. Rev.D}, 1987, vol 33, p. 1191.
\bibitem{Berry}
Berry, M., {\sl Proc.R.Soc.Lond., Ser. A}, 1984, vol. 392, p.45.
\bibitem{aharon}
Aharonov, Y. and Anandan, J.,
{\sl Phys. Rev. Lett.}, 1987, vol. 58, p. 1593.

\end{thebibliography}
\end{document}